\documentclass[letterpaper,12pt]{JHEP}
\usepackage{amsmath,amssymb}
\usepackage{epsfig}
\usepackage{cite}
\usepackage[english]{babel}

\newcommand{\dd}{{\rm d}}

\newcommand{\eq}{\begin{equation}}
\newcommand{\feq}{\end{equation}}
\newcommand{\eqn}{\begin{eqnarray}}
\newcommand{\feqn}{\end{eqnarray}}
\newcommand{\arr}{\begin{eqnarray*}}
\newcommand{\farr}{\end{eqnarray*}}

\newcommand{\M}{{\cal M}}

\font\mybb=msbm10 at 12pt
\def\bb#1{\hbox{\mybb#1}}

\def\bR {\bb{R}}

\def\bC {\bb{C}}

\def\la{\lambda}

\def\si{\sigma}

\def\La{\Lambda}

\newcommand{\lcc}[1]{\tilde{#1}}

\title{Chern-Simons formulation of three-dimensional gravity with torsion and nonmetricity}
\author{Sergio L.~Cacciatori \\
Dipartimento di Matematica dell'Universit\`a di Milano \\
Via Saldini 50, I-20133 Milano and \\
INFN, Sezione di Milano, Via Celoria 16, I-20133 Milano. \\
E-mail: \email{cacciatori@mi.infn.it}}
\author{Marco M.~Caldarelli, Alex Giacomini, Dietmar Klemm and Diego S.~Mansi \\
Dipartimento di Fisica dell'Universit\`a di Milano \\
Via Celoria 16, I-20133 Milano and \\
INFN, Sezione di Milano, Via Celoria 16, I-20133 Milano. \\
E-mail: \email{marco.caldarelli@mi.infn.it,
               alex.giacomini@mi.infn.it,
               dietmar.klemm@mi.infn.it,
               diego.mansi@mi.infn.it}}
\preprint{IFUM-839-FT \\
hep-th/0507200}

\abstract{We consider various models of three-dimensional gravity with torsion or
nonmetricity (metric affine gravity), and show that they can be written as Chern-Simons
theories with suitable gauge groups. Using the groups ISO$(2,1)$, SL$(2,\bC)$ or SL$(2,\bR)$
$\times$ SL$(2,\bR)$, and the fact that they admit two independent coupling constants,
we obtain the Mielke-Baekler model for zero, positive or negative effective
cosmological constant respectively. Choosing SO$(3,2)$ as gauge group, one gets a
generalization of conformal gravity that has zero torsion and only the trace part
of the nonmetricity. This characterizes a Weyl structure. Finally, we present a new topological
model of metric affine gravity in three dimensions arising from an SL$(4,\bR)$
Chern-Simons theory.
}


\dedicated{Dedicated to the memory of Antonio Meucci}

\begin{document}

\section{Introduction}
\label{intro}

General relativity in four spacetime dimensions is a notoriously difficult
theory, already at the classical and in particular at the quantum level.
This is one of the main reasons why people are interested in simpler
models that nevertheless retain almost all of the essential features
of four-dimensional general relativity. One such model is pure gravity
in 2+1 dimensions, with or without cosmological constant. This theory
has been studied extensively in the past, in particular by Deser, Jackiw
and `t Hooft \cite{Deser:1983tn,Deser:1983dr}.
The most famous example where we learned something on general relativity
by considering a simpler toy model is perhaps the BTZ black hole \cite{Banados:1992wn},
whose study revealed a lot on the quantum structure and the statistical mechanics
of black holes (for a review cf.~\cite{Carlip:1998qw}).

Major progress in 2+1 dimensional gravity came when Ach\'{u}carro and Townsend
\cite{Achucarro:1987vz} and Witten \cite{Witten:1988hc} showed that
these systems can be written as Chern-Simons (CS) theories, with gauge group
ISO$(2,1)$, SL$(2,\bC)$ or SL$(2,\bR)$ $\times$ SL$(2,\bR)$ for zero, positive or negative
cosmological constant respectively. In trying to write down a CS action for
the Poincar\'{e} group, one encounters the problem that ISO$(2,1)$ is not semisimple,
and therefore the Killing form is degenerate. As noted by Witten \cite{Witten:1988hc},
the Poincar\'{e} algebra admits nevertheless a nondegenerate, Ad-invariant bilinear form
given by
\begin{equation}
\langle J_a\,, P_b\rangle = \eta_{ab}\,, \qquad \langle J_a\,, J_b\rangle = \lambda\,\eta_{ab}\,, \qquad
\langle P_a\,, P_b\rangle = 0\,, \label{quadform}
\end{equation}
where $J_a$ and $P_a$ denote the Lorentz and translation generators, and $\lambda$ is
an arbitrary real constant. Mathematically, the existence of an Ad-invariant, nondegenerate
quadratic form on the Poincar\'{e} algebra follows from the fact that iso$(2,1)$
is the double extension of a reductive Lie algebra (in this case the trivial algebra):
Let $\cal A$ be a reductive Lie algebra, i.~e.~, a direct sum
of a semisimple and an abelian algebra. $\cal A$ admits an invariant nondegenerate
bilinear form $\Omega_{ij}$, whose restriction to the semisimple part is simply given
by the Killing form, and the restriction to the abelian subalgebra is proportional to
the identity. The generators $\tau_i$ of $\cal A$ satisfy
\begin{displaymath}
[\tau_i, \tau_j] = {f_{ij}}^k \tau_k\,.
\end{displaymath}
The double extension of $\cal A$ is obtained by adding the new generators $H_a$
and $H^*_{\bar b}$ such that
\begin{eqnarray}
[\tau_i, \tau_j] &=& {f_{ij}}^k \tau_k + {h_{ij}}^{\bar a}H^*_{\bar a}\,, \nonumber \\
{[}H_a, \tau_i] &=& {h_{a i}}^j \tau_j\,, \nonumber \\
{[}H_a, H_b] &=& {g_{ab}}^c H_c\,, \label{doubleext} \\
{[}H_a, H^*_{\bar b}] &=& {g_{a\bar b}}^{\bar c} H^*_{\bar c}\,,
                                      \nonumber \\
{[}\tau_i, H^*_{\bar a}] &=& [H^*_{\bar a}, H^*_{\bar b}] = 0\,, \nonumber
\end{eqnarray}
where ${h_{a i}}^j \Omega_{jk} = {h_{ik}}^{\bar a} \delta_{{\bar a} a}$.
If furthermore ${g_{ab}}^c = {g_{a\bar b}}^{\bar c}$, there exists an Ad-invariant, nondegenerate
quadratic form on the double extension of $\cal A$, given by \cite{Figueroa-O'Farrill:1994yf}
\begin{equation}
\Omega_{IJ} = \left(\begin{array}{ccc} \Omega_{ij} & 0 & 0 \\
                                  0 & \lambda_{ab} & \delta_{a \bar b} \\
                                  0 & \delta_{{\bar a} b} & 0 \end{array} \right)\,, \label{quadformgen}
\end{equation}
where $I = i,a,\bar a$, and $\lambda_{ab}$ denotes any invariant quadratic form on the algebra
generated by the $H_a$.
If the algebra $\cal A$ is trivial (no generators $\tau_i$), (\ref{doubleext}) has
exactly the structure of the Poincar\'{e} algebra
\begin{equation}
[J_a, J_b] = {\epsilon_{ab}}^c J_c\,, \qquad [J_a, P_b] = {\epsilon_{ab}}^c P_c\,, \qquad
[P_a, P_b] = 0\,,
\end{equation}
if we identify the generators $H_a$ with $J_a$ and $H^*_{\bar b}$ with
$P_b$. The invariant quadratic form (\ref{quadformgen}) reduces then to (\ref{quadform}).
Retaining a nonvanishing $\lambda$ in a CS formulation of three-dimensional gravity leads
to the inclusion of a gravitational Chern-Simons action (i.~e.~, a CS term for the spin
connection) \cite{Witten:1988hc}. This does not change the classical equations of motion,
but leads to modifications at the quantum level \cite{Witten:1988hc}.\\
One can now try to depart from pure gravity, rendering thus the model
less trivial, while maintaining at the same time its integrability.
A possible way to introduce additional structure is to permit nonvanishing torsion and/or
nonmetricity (metric affine gravity) \cite{Hehl:1994ue}.
We would like to do this in such a way that the resulting model
can still be written as a CS theory for some gauge group. There are several reasons that
motivate the introduction of torsion or nonmetricity. Let us mention here only a few of them.
For a more detailed account we refer to \cite{Hehl:1994ue}.
First of all, nonmetricity is a measure
for the violation of local Lorentz invariance \cite{Hehl:1994ue}, which has become fashionable
during the last years. Second, the geometrical concepts of nonmetricity and torsion have
applications in the theory of defects in crystals, where they are interpreted as densities
of point defects and line defects (dislocations) respectively, cf.~\cite{Kroener:1990}
and references therein. Finally, nonmetric connections or connections with torsion are
interesting from a mathematical point of view. For example, a torsionless connection that
has only the trace part of the nonmetricity characterizes a so-called Weyl structure.
If, moreover, the symmetric part of the Ricci tensor is proportional to the metric, one has
an Einstein-Weyl structure (cf.~e.~g.~\cite{Dunajski:2000rf}). Einstein-Weyl manifolds
represent the analogue of Einstein spaces in Weyl geometry,
and are less trivial than the latter, which have necessarily constant curvature in three
dimensions. Einstein-Weyl structures are interesting also due to their relationship
to certain integrable systems, like the SU$(\infty)$ Toda \cite{Ward:1990qt} or the
dispersionless Kadomtsev-Petviashvili equation \cite{Dunajski:2000rf}.\\
In this paper, we consider various models of three-dimensional metric-affine gravity
and show that they can be written as CS theories. This is accomplished either by
using gauge groups larger than ISO$(2,1)$, SL$(2,\bC)$ or SL$(2,\bR)$ $\times$ SL$(2,\bR)$,
or by using the fact that these groups admit two independent coupling constants, as was
explained above for the case of the Poincar\'{e} group.\\
The remainder of our paper is organized as follows: In the next section, we briefly
summarize the basic notions of metric affine gravity. In section 3, we show that the
Mielke-Baekler model, which is characterized by nonvanishing torsion and zero nonmetricity,
can be written as a CS theory for arbitrary values of the effective cosmological
constant. In section 4, a CS action for the conformal group SO$(3,2)$
is considered, and it is shown that this leads to a generalization of conformal gravity
with a Weyl connection. Finally, in section 5, we propose a topological model of metric
affine gravity based on an SL$(4,\bR)$ CS theory and discuss some of its solutions. In the
last section we summarize the results and draw some conclusions.

\section{Metric affine gravity}
\label{mag}

In order to render this paper self-contained, we summarize briefly the basic notions of metric
affine gravity. For a detailed review see \cite{Hehl:1994ue}.

The standard geometric setup of Einstein's general relativity is a differential ma\-ni\-fold $\M$,
of dimension $D$, endowed with a metric $g$ and a Levi-Civita
connection $\lcc\nabla$, which is uniquely determined by the requirements of metricity
($\lcc\nabla g=0$) and vanishing torsion. 
This structure is known as a semi-Riemannian space $(\M,g)$.

One can now consider more complicated non-Riemannian geometries, where
a new generic connection $\nabla$ is introduced on $T\M$ which is, in general, independent
of the metric. In this way one defines a new mathematical structure called a
\emph{metric-affine} space $(\M,g,\nabla)$.

One can measure the deviation from the standard geometric setup by computing the difference
$(\nabla-\lcc\nabla)v$ between the action of the two connections on a vector field $v$
defined on $T\M$. To be more specific one can choose a chart, so that the action of the
connection is described by its coefficients,\footnote{The connection coefficients for the Levi-Civita
connection are called Christoffel symbols and are denoted by $\tilde\Gamma^\nu{}_{\mu\la}$.}
\eq
\nabla_{\mu} v^{\nu} = \partial_{\mu} v^{\nu} + \Gamma^{\nu}{}_{\mu\lambda}v^{\lambda}\,,
\feq
and the deviation can be written as
\eq
(\nabla_{\mu}-\lcc{\nabla}_{\mu})v^{\nu} = N^{\nu}{}_{\mu\lambda}v^{\lambda}\,.
\feq
The tensor
\eq
N^{\lambda}{}_{\mu\nu} = \Gamma^{\lambda}{}_{\mu\nu} - \lcc\Gamma{}^{\lambda}{}_{\mu\nu}
                         \label{distortion}
\feq
is called \emph{distortion} and measures the deviation of $\nabla$ from the Levi-Civita
connection. This object can be decomposed in different parts,
depending essentially on two quantities: the \emph{torsion} and the \emph{nonmetricity}.

The torsion tensor $T^a$ is defined by the first Cartan structure equation
\eq
T^{a} \equiv \dd e^{a}+\omega^{a}{}_{b}\wedge e^{b}\,,
\feq
where $\omega^a{}_b$ is the spin connection acting on (flat) tangent space
indices $a,b,\ldots$, and $e^a$ denotes the vielbein satisfying
${e^a}_{\mu} {e^b}_{\nu} g^{\mu\nu} = \eta^{ab}$, with $\eta^{ab}$ the flat
Minkowski metric. A priori, $\omega^a{}_b$
is independent of the connection coefficients $\Gamma^{\lambda}{}_{\mu\nu}$.
Both objects become dependent of each other by the tetrad postulate
\begin{equation}
\nabla_{\mu} {e^a}_{\nu} = 0\,, \label{tetrpost}
\end{equation}
implying
\eq
\omega^a_{\mu\,b} = e^{a}{}_{\lambda}\Gamma^{\lambda}{}_{\mu\rho}\,e_{b}{}^{\rho} -
                    e_{b}{}^{\lambda}\partial_{\mu} e^{a}{}_{\lambda}\,,
\feq
so that the spin connection $\omega^a_{\mu\,b}$ is the gauge transform of
$\Gamma^{\lambda}{}_{\mu\rho}$ with transformation matrix $e^{a}{}_{\lambda}$.

For nonvanishing torsion, the connection coefficients are no more symmetric in
their lower indices, as can be seen from
\eq
0 = 2\nabla_{[\mu}{e^a}_{\nu]} = \left(\partial_{\mu}e^{a}{}_{\nu} - \partial_{\nu}e^{a}{}_{\mu}
    + \omega^a_{\mu\,b} e^{b}{}_{\nu} - \omega^a_{\nu\,b} e^{b}{}_{\mu}\right)
    - 2\Gamma^{\lambda}{}_{[\mu\nu]}e^{a}{}_{\lambda}\,,
\feq
which yields
\eq
T^{\lambda}{}_{\mu\nu} \equiv e_{a}{}^{\lambda}T^{a}{}_{\mu\nu} = 2\Gamma^{\lambda}{}_{[\mu\nu]}\,,
\feq
or, equivalently,
\begin{equation}
T^{\lambda}{}_{\mu\nu} = 2N^{\lambda}{}_{[\mu\nu]}\,, \label{TN}
\end{equation}
since the Levi-Civita connection has zero torsion, $\lcc\Gamma^{\lambda}{}_{[\mu\nu]}=0$.

The nonmetricity $Q$ is a tensor which measures the failure of the metric to be covariantly constant,
\eq
Q_{\lambda\mu\nu} \equiv -\nabla_{\mu}g_{\nu\lambda}\,.
\feq
Using $\lcc\nabla g = 0$ and the definition (\ref{distortion}), one gets
\eq
Q_{\lambda\mu\nu} = N_{\lambda\mu\nu} + N_{\nu\mu\lambda}\,, \label{QN}
\feq
where $N_{\lambda\mu\nu} = g_{\lambda\sigma} N^{\sigma}{}_{\mu\nu}$. For nonzero nonmetricity,
the spin connection $\omega^{ab}$ is no more antisymmetric in $a,b$: By computing the covariant
derivative $\nabla_{\mu} \eta^{ab}$ one obtains $Q_{\mu}^{ab} = 2\omega_{\mu}^{(ab)}$,
with $Q_{\mu}^{ab} \equiv e^{a\rho} e^{b\lambda} Q_{\rho\mu\lambda}$. This means that the spin
connection takes values in gl$(D,\bR)$ instead of the Lorentz algebra so$(D-1, 1)$.

Notice that in presence of nonmetricity, the scalar product of two vectors $u$,
$v$ can change when $u$, $v$ are parallel transported along a curve. Let $t$ be the tangent vector
of an infinitesimal curve $c$. The variation of the scalar product is then given by
\eq
\delta g(u, v) = \nabla_t (g_{\mu\nu} u^{\mu} v^{\nu}) =
                 -Q_{\mu\lambda\nu} t^{\lambda} u^{\mu} v^{\nu}\,.
\feq
Physically, this states that if we enlarge the Lorentz group, the interval is not any longer an
invariant and in fact, for generic nonmetricity, the very concept of light cone is lost.

The two tensors $T$ and $Q$ uniquely determine the distortion and, as a result, the connection.
This can also be seen by counting the degrees of freedom: the distortion is a generic tensor with three indices,
so it has $D^3$ independent components. The torsion and the non-metricity, due to their symmetry properties,
have respectively $D^{2}(D-1)/2$ and $D^{2}(D+1)/2$ independent components; their sum gives precisely the expected
number of degrees of freedom. To obtain the distortion in terms of torsion and nonmetricity one has to solve
the equations (\ref{TN}) and (\ref{QN}).
Considering all possible permutations one obtains
\eq
N_{\lambda\mu\nu} = \frac12\left(T_{\nu\lambda\mu} - T_{\lambda\nu\mu} - T_{\mu\nu\lambda}\right)
+ \frac12\left(Q_{\lambda\mu\nu} + Q_{\lambda\nu\mu} - Q_{\mu\lambda\nu}\right)\,,
\label{N}
\feq
which is the expected decomposition of the distortion. The Levi-Civita connection
is obtained setting $T^a = 0$ and $Q_{ab}=0$. The combination
\eq
K_{\nu\lambda\mu} = \frac12\left(T_{\nu\lambda\mu} - T_{\lambda\nu\mu} - T_{\mu\nu\lambda}\right)\,,
                    \label{contorsion}
\feq
which is antisymmetric in the first two indices, is also called contorsion.

Note that in metric affine gravity, the local symmetry group is the affine group
A$(D,\bR)$ $\cong$ GL$(D,\bR)$ $\ltimes$ $\bR^D$ instead of the Poincar\'{e} group
ISO$(D-1,1)$. The associated gauge fields are $\omega^{ab}$ and $e^a$.
In what follows, we shall specialize to the case $D=3$.

\section{The Mielke-Baekler model as a Chern-Simons theory}
\label{MB}

Let us first consider the case of Riemann-Cartan spacetimes,
characterized by vanishing nonmetricity, but nonzero
torsion. A simple three-dimensional
model that yields nonvanishing torsion was proposed by Mielke and Baekler
(MB) \cite{Mielke:1991nn} and further analyzed by Baekler, Mielke and
Hehl \cite{Baekler:1992ab}. The action reads \cite{Mielke:1991nn}
\begin{equation}
I = a I_1 + \Lambda I_2 + \alpha_3 I_3 + \alpha_4 I_4\,, \label{MBaction}
\end{equation}
where $a$, $\Lambda$, $\alpha_3$ and $\alpha_4$ are constants,
\begin{eqnarray}
I_1 &=& 2\int e_a \wedge R^a\,, \nonumber \\
I_2 &=& -\frac 13 \int \epsilon_{abc} e^a \wedge e^b \wedge e^c\,, \nonumber \\
I_3 &=& \int \omega_a \wedge d\omega^a + \frac 13 \epsilon_{abc}
        \omega^a \wedge \omega^b \wedge \omega^c\,, \nonumber \\
I_4 &=& \int e_a \wedge T^a\,, \nonumber
\end{eqnarray}
and
\begin{eqnarray}
R^a &=& d\omega^a + \frac 12 {\epsilon^a}_{bc}\,\omega^b \wedge \omega^c\,, \nonumber \\
T^a &=& de^a + {\epsilon^a}_{bc}\,\omega^b \wedge e^c\,, \nonumber
\end{eqnarray}
denote the curvature and torsion two-forms respectively.
$\omega^a$ is defined by $\omega^a = \frac 12 \epsilon^{abc}\omega_{bc}$ with
$\epsilon_{012} = 1$.
$I_1$ yields the Einstein-Hilbert
action, $I_2$ a cosmological constant, $I_3$ is a Chern-Simons term for the connection,
and $I_4$ represents a translational Chern-Simons term.
Note that, in order to obtain the topologically massive gravity of Deser, Jackiw and
Templeton (DJT) \cite{Deser:1981wh} from (\ref{MBaction}), one has to add a Lagrange multiplier
term that ensures vanishing torsion.
The field equations following from (\ref{MBaction}) take the form
\begin{eqnarray}
2a R^a - \Lambda {\epsilon^a}_{bc}\,e^b \wedge e^c + 2\alpha_4 T^a &=& 0\,, \nonumber \\
2a T^a + 2\alpha_3 R^a + \alpha_4 {\epsilon^a}_{bc}\,e^b \wedge e^c &=& 0\,. \nonumber
\end{eqnarray}
In what follows, we assume $\alpha_3\alpha_4 - a^2 \neq 0$\footnote{For
$\alpha_3\alpha_4 - a^2 = 0$ the theory becomes singular \cite{Baekler:1992ab}.}.
Then the equations of motion can be rewritten as
\begin{equation}
2T^a = A{\epsilon^a}_{bc}\,e^b \wedge e^c\,, \qquad
2R^a = B{\epsilon^a}_{bc}\,e^b \wedge e^c\,, \label{eomMB}
\end{equation}
where
\begin{displaymath}
A = \frac{\alpha_3\Lambda + \alpha_4 a}{\alpha_3\alpha_4 - a^2}\,, \qquad
B = -\frac{a\Lambda+\alpha_4^2}{\alpha_3\alpha_4 - a^2}\,.
\end{displaymath}
Thus, the field configurations are characterized by constant curvature and
constant torsion. The curvature $R^a$ of a Riemann-Cartan
spacetime can be expressed in terms of its Riemannian part ${\tilde R}^a$ and
the contorsion one-form $K^a$ by
\begin{equation}
R^a = {\tilde R}^a - dK^a - {\epsilon^a}_{bc}\,\omega^b \wedge K^c - \frac 12
      {\epsilon^a}_{bc}\,K^b \wedge K^c\,, \label{RK}
\end{equation}
where $K^a_{\mu} = \frac 12 {\epsilon^a}_{bc}\,e^{b\beta} e^{c\gamma} K_{\beta\gamma\mu}$,
and $K_{\beta\gamma\mu}$ denotes the contorsion tensor given by (\ref{contorsion}).
Using the equations of motion (\ref{eomMB}) in (\ref{RK}), one gets for the
Riemannian part
\begin{equation}
2{\tilde R}^a = \Lambda_{\mathrm{eff}}{\epsilon^a}_{bc}\,e^b \wedge e^c\,,
\end{equation}
with the effective cosmological constant
\begin{displaymath}
\Lambda_{\mathrm{eff}} = B - \frac{A^2}4\,.
\end{displaymath}
This means that the metric is given by the (anti-)de~Sitter or Minkowski solution,
depending on whether $\Lambda_{\mathrm{eff}}$ is negative, positive or zero.
It is interesting to note that $\Lambda_{\mathrm{eff}}$ can be nonvanishing
even if the bare cosmological constant $\Lambda$ is zero \cite{Baekler:1992ab}.
In this simple model, dark energy (i.~e.~, $\Lambda_{\mathrm{eff}}$) would then be
generated by the translational Chern-Simons term $I_4$.

In \cite{Blagojevic:2003vn} it was shown that for $\Lambda_{\mathrm{eff}} < 0$, the
Mielke-Baekler model (\ref{MBaction}) can be written as a sum of two SL$(2,\bR)$
Chern-Simons theories. We will now shew that this can be generalized to the case
of arbitrary effective cosmological constant. For positive $\Lambda_{\mathrm{eff}}$, the
action $I$ becomes a sum of two SL$(2,\bC)$ Chern-Simons theories with complex coupling
constants, whereas for vanishing $\Lambda_{\mathrm{eff}}$, $I$ can be written as
CS theory for the Poincar\'{e} group.

To start with, we briefly summarize the results of \cite{Blagojevic:2003vn}. For
$\Lambda_{\mathrm{eff}} < 0$ the geometry is locally AdS$_3$, which has the
isometry group SO$(2,2)$ $\cong$ SL$(2,\bR) \times$ SL$(2,\bR)$, so if the MB model
is equivalent to a Chern-Simons theory, one expects a gauge group SO$(2,2)$.
Indeed, if one defines the SL$(2,\bR)$ connections
\begin{displaymath}
A^a = \omega^a + q\, e^a\,, \qquad {\tilde A}^a = \omega^a + \tilde q\, e^a\,,
\end{displaymath}
then the SL$(2,\bR) \times$ SL$(2,\bR)$ Chern-Simons action\footnote{In (\ref{CSMBAdS}),
$\langle\tau_a\,,\tau_b\rangle = 2\,\mathrm{Tr}\,(\tau_a\tau_b) = \eta_{ab}$, and the SL$(2,\bR)$
generators $\tau_a$ satisfy $[\tau_a, \tau_b] = {\epsilon_{ab}}^c\tau_c$.}
\begin{equation}
I_{CS} = \frac t{8\pi} \int \langle A \wedge dA + \frac 23 A \wedge A \wedge A\rangle + \frac{\tilde t}{8\pi}
         \int \langle\tilde A \wedge d\tilde A + \frac 23 \tilde A \wedge \tilde A \wedge \tilde A\rangle
         \label{CSMBAdS}
\end{equation}
coincides (up to boundary terms) with $I$ in (\ref{MBaction}), if the parameters
$q, \tilde q$ and the coupling constants $t, \tilde t$ are given by
\begin{equation}
q = -\frac A2 + \sqrt{-\Lambda_{\mathrm{eff}}}\,, \qquad
\tilde q = -\frac A2 - \sqrt{-\Lambda_{\mathrm{eff}}}\,
\end{equation}
and
\begin{equation}
\frac t{2\pi} = 2\alpha_3 + \frac{2a + \alpha_3 A}{\sqrt{-\Lambda_{\mathrm{eff}}}}\,, \qquad
\frac{\tilde t}{2\pi} = 2\alpha_3 - \frac{2a + \alpha_3 A}{\sqrt{-\Lambda_{\mathrm{eff}}}}\,.
\end{equation}
We see that $q, \tilde q$, and thus the connections $A^a, {\tilde A}^a$ are real for negative
$\Lambda_{\mathrm{eff}}$. The coupling constants $t, \tilde t$ are also real, but in general
different from each other due to the presence of $I_3$.

For $\Lambda_{\mathrm{eff}} > 0$, $q$ and $\tilde q$ become complex, with $\tilde q = \bar q$
and thus ${\tilde A}^a = {\bar A}^a$. As the connections are no more real, we must consider the
complexification SL$(2,\bC)$ of SL$(2,\bR)$. Then (\ref{CSMBAdS}) becomes a sum of two
SL$(2,\bC)$ Chern-Simons actions, with complex coupling constants $t, \tilde t$,
where $\tilde t = \bar t$. Again, (\ref{CSMBAdS}) is equal (modulo boundary terms)
to the Mielke-Baekler action (\ref{MBaction}). This makes of course sense, since the isometry
group of three-dimensional de~Sitter space is SO$(3,1)$ $\cong$ SL$(2,\bC)$. The usual
CS formulation of dS$_3$ gravity \cite{Witten:1988hc} is recovered for $\alpha_3 = \alpha_4 = 0$.

The real part of $t$, i.~e.~, up to prefactors, $\alpha_3$, is subject to a topological
quantization condition coming from the maximal compact subgroup SU$(2)$ of
SL$(2,\bC)$ \cite{Witten:1989ip}. As $\tilde t = \bar t$, the action (\ref{CSMBAdS})
leads to a unitary quantum field theory \cite{Witten:1989ip}.

Finally, we come to the case of vanishing $\Lambda_{\mathrm{eff}}$. The condition
$B - A^2/4 = 0$ implies that $\Lambda$ can be expressed in terms of the other parameters
according to
\begin{equation}
\Lambda = \frac{2a^3 - 3a \alpha_3\alpha_4 \pm 2\,(a^2 - \alpha_3\alpha_4)^{\frac 32}}{\alpha_3^2}\,.
          \label{Lambda}
\end{equation}
As we want $\Lambda$ to be real, we assume $a^2 - \alpha_3\alpha_4 > 0$. Let us consider
the CS action
\begin{equation}
I_{CS} = \frac k{4\pi}\int \langle A \wedge dA + \frac 23 A \wedge A \wedge A\rangle\,,
         \label{CSMBMink}
\end{equation}
where $A$ denotes an iso$(2,1)$ valued connection, and the quadratic form on the
Poincar\'{e} algebra is given by (\ref{quadform}).
According to what was said in the introduction, this (nondegenerate) bilinear
form is Ad-invariant for any value of the parameter $\lambda$.
If we decompose the connection as
\begin{equation}
A = e^a P_a + (\omega^a + \gamma e^a) J_a\,,
\end{equation}
then the CS action (\ref{CSMBMink}) coincides, up to boundary terms, with
(\ref{MBaction}) (where now $\Lambda$ is not independent, but determined by (\ref{Lambda})),
if the constants $k$, $\lambda$ and $\gamma$ are chosen as
\begin{equation}
\frac k{4\pi} = \mp\sqrt{a^2 - \alpha_3\alpha_4}\,, \qquad
\lambda = \mp\frac{\alpha_3}{\sqrt{a^2 - \alpha_3\alpha_4}}\,, \qquad
\gamma = \frac{a \pm\sqrt{a^2 - \alpha_3\alpha_4}}{\alpha_3}\,.
\end{equation}
In conclusion, we have shown that the Mielke-Baekler model can be written as
a Chern-Simons theory for any value of the effective cosmological constant
$\Lambda_{\mathrm{eff}}$, whose sign determines the gauge group. This was
accomplished by a nonstandard decomposition of the CS connection in terms of
the dreibein and the spin connection, and by using the fact that the considered
gauge groups admit two independent coupling constants. As the CS connection is flat,
and thus entirely determined by holonomies, there are no propagating local degrees
of freedom; hence there cannot be any gravitons in the MB model,
contrary to the claim in \cite{Baekler:1992ab}.

It would be interesting to study the asymptotic dynamics of the Mielke-Baekler
model in the case $\Lambda_{\mathrm{eff}} < 0$, where the spacetime is locally
AdS$_3$.
According to the AdS/CFT correspondence \cite{Aharony:1999ti}, (\ref{MBaction})
should then be equivalent to a two-dimensional conformal field theory
on the boundary of AdS$_3$,
where the bulk fields $e^a$ and $\omega^a$ are sources for the CFT
energy-momentum current and spin current respectively.
It was claimed in \cite{Blagojevic:2003vn} that in general the putative CFT
has two different central charges. (Unlike the case $\alpha_3 = \alpha_4 = 0$,
$a = 1/16\pi G$, $\Lambda = -1/l^2$, where $c_L = c_R = 3l/2G$ \cite{Brown:1986nw}).
It would be interesting to compute these central charges explicitely, and to see
whether the entropy of the Riemann-Cartan black hole \cite{Garcia:2003nm}
(which represents a generalization of the BTZ black hole with torsion) can be
reproduced by counting CFT states using the Cardy formula.
Similar to \cite{Coussaert:1995zp},
one expects the action (\ref{CSMBAdS}) to reduce to a sum of two chiral
WZNW actions on the conformal spacetime boundary. For $\alpha_3 =\alpha_4 = 0$,
these two chiral actions combine into a single nonchiral WZNW model \cite{Coussaert:1995zp}.
As the two SL$(2,\bR)$ CS actions in (\ref{CSMBAdS}) have different coupling
constants, it might be that in the general case this is no more true, and one is left
with a sum of two chiral WZNW models that have different central charges. It remains to
be seen how this reduction works in detail.

\section{Weyl structures from Chern-Simons theory}

In this section we will show how to get Weyl structures, which are
characterized by torsion-free connections that involve only the trace part
of the nonmetricity, from Chern-Simons theory. To start with, let us
consider conformal gravity in three dimensions, defined by the
action \cite{Deser:1981wh}
\begin{equation}
I = \int\left({\omega^a}_b \wedge d{\omega^b}_a + \frac 23
        {\omega^a}_b \wedge {\omega^b}_c \wedge {\omega^c}_a\right)\,.
        \label{confgrav}
\end{equation}
Here, $\omega$ denotes an so$(2,1)$ valued (and hence metric) connection,
which is not a fundamental variable, but is considered as a function
of the dreibein, as is required by vanishing torsion. Therefore,
variation of (\ref{confgrav}) leads to third order differential equations,
namely \cite{Deser:1981wh}
\begin{equation}
C^{\mu\nu} \equiv \frac 1{\sqrt{-g}} \epsilon^{\mu \alpha \beta}
\nabla _{\alpha} {L^{\nu}}_{\beta} = 0\,, \label{conformal}
\end{equation}
where $L_{\mu\nu}$ denotes the Schouten tensor defined by
\begin{equation}
L_{\mu \nu} = R_{\mu \nu} - \frac 14 R\,g_{\mu\nu}\,.
\end{equation}
$C^{\mu\nu}$ is known as Cotton-York
tensor\footnote{$C_{\mu\rho\nu}= \nabla_{\mu} L_{\nu\rho} - \nabla_{\rho} L_{\nu\mu}$
is called the Cotton two-form. See \cite{Garcia:2003bw} for a nice review.}.
It has zero trace, reflecting the
conformal invariance of (\ref{confgrav}). In three dimensions, the Cotton-York
tensor takes the role of the Weyl tensor (which is identically zero in 3d).
$C_{\mu\nu}$ vanishes if and only if spacetime is conformally flat \cite{Eisenhart}.
(\ref{confgrav}) is sometimes called the gravitational Chern-Simons
action. Its supersymmetric extension was obtained in \cite{Deser:1982sw}.
The dimensional reduction of the action (\ref{confgrav}), studied in
\cite{Guralnik:2003we}, has recently been shown to describe a subsector of BPS
solutions to gauged supergravity in four dimensions \cite{Cacciatori:2004rt}.

Originally, the gravitational Chern-Simons action was introduced by Deser, Jackiw
and Templeton in order to render three-dimensional Einstein gravity nontrivial: If
one adds (\ref{confgrav}) to the Einstein-Hilbert action, the theory acquires a
propagating, massive, spin 2 degree of freedom \cite{Deser:1981wh}.

Horne and Witten showed that conformal gravity in three dimensions with action
(\ref{confgrav}) can be written as a Chern-Simons theory for the conformal
group SO$(3,2)$ \cite{Horne:1988jf}. To this end, they decomposed the
SO$(3,2)$ connection $A$ according to
\begin{equation}
A_{\mu} = {e^a}_{\mu} P_a - \frac 12\omega^{ab}_{\mu} J_{ab} + {\lambda^a}_{\mu} K_a
          + \phi_{\mu} D\,,
\end{equation} 
where $P_a, J_{ab}, K_a, D$ denote respectively the generators of translations,
Lorentz transformations, special conformal transformations and dilations.
The Chern-Simons action for $A$ leads then to the equations of motion \cite{Horne:1988jf}
\begin{eqnarray}
&& de^a + {\omega^a}_b \wedge e^b - \phi \wedge e^a = 0\,, \label{e} \\
&& d\omega^{ab} + {\omega^a}_c \wedge \omega^{cb} - e^a \wedge \lambda^b + e^b \wedge \lambda^a = 0\,,
   \label{omega} \\
&& d\lambda^a + {\omega^a}_b \wedge \lambda^b + \phi \wedge \lambda^a = 0\,, \label{lambda} \\
&& d\phi + e^a \wedge \lambda_a = 0\,. \label{phi}
\end{eqnarray}
The generator of an infinitesimal gauge transformation is
\begin{displaymath}
u = \rho^a P_a - \frac 12 \tau^{ab} J_{ab} + \sigma^a K_a + \gamma D\,.
\end{displaymath}
The transformation law $\delta A = -d u - [A, u]$ leads then
to
\begin{eqnarray}
\delta e^a &=& -d\rho^a - {\omega^a}_b \rho^b + e^b {\tau^a}_b - e^a \gamma + \phi \rho^a\,, \nonumber \\
\delta \omega^{ab} &=& -d\tau^{ab} - {\omega^a}_c \tau^{cb} + {\omega^b}_c \tau^{ca} +
                       e^a \sigma^b - e^b \sigma^a + \lambda^a \rho^b - \lambda^b \rho^a\,, \nonumber \\
\delta \lambda^a &=& -d\sigma^a - {\omega^a}_b \sigma^b + \lambda^b {\tau^a}_b + \lambda^a \gamma
                     - \phi \sigma^a\,, \nonumber \\
\delta \phi &=& -d\gamma - e^a \sigma_a + \lambda^a \rho_a\,. \label{transfcomp}
\end{eqnarray}
Horne and Witten noticed that when the vielbein ${e^a}_{\mu}$ is invertible,
the $\sigma^a$ gauge invariance is precisely sufficient to set $\phi = 0$.
With the gauge choice $\phi = 0$, the equations of motion simplify considerably.
(\ref{e}) implies then that the torsion vanishes. If we define
$\lambda_{\mu\nu} = e_{a\mu} {\lambda^a}_{\nu}$, eq.~(\ref{phi}) means that
$\lambda_{\mu\nu}$ is symmetric, whereas (\ref{omega}) leads to
\begin{equation}
\lambda_{\mu\nu} = R_{\mu\nu} - \frac R4\,g_{\mu\nu} = L_{\mu\nu}\,,
\end{equation}
so that $\lambda_{\mu\nu}$ represents the Schouten tensor. It is interesting
to note that in this context, the Schouten tensor, which physically corresponds to
a curvature, is at the same time a connection, namely the gauge field of special conformal
transformations. Eq.~(\ref{omega}) is then precisely the expression for the Riemann
curvature tensor in terms of the Schouten tensor,
\begin{displaymath}
R_{\mu\nu\rho\sigma} = g_{\mu\rho} L_{\nu\sigma} + g_{\nu\sigma} L_{\mu\rho} -
                       g_{\mu\sigma} L_{\nu\rho} - g_{\nu\rho} L_{\mu\sigma}\,,
\end{displaymath}
valid in three dimensions. Finally, one has to interpret (\ref{lambda}). To this
end, one defines the connection coefficients ${\Gamma^{\mu}}_{\nu\rho}$ by
requiring (\ref{tetrpost}), which implies
\begin{displaymath}
{\Gamma^{\mu}}_{\nu\rho} = {e_a}^{\mu} \omega^a_{\nu b}\,{e^b}_{\rho} + {e_a}^{\mu}\partial_{\nu}
                           {e^a}_{\rho}\,.
\end{displaymath}
Eq.~(\ref{lambda}) is then equivalent to
\begin{equation}
\nabla_{\mu} L_{\nu\rho} - \nabla_{\rho} L_{\nu\mu} = 0\,,
\end{equation}
which coincides with the equation of motion (\ref{conformal}) following from the action
(\ref{confgrav}). In the gauge $\phi = 0$, the gauge theory of the conformal group
with Chern-Simons action is therefore equivalent to conformal gravity.

We can now ask what happens if one does not set $\phi = 0$. In this case it is
convenient to define a generalized connection $\hat{\omega}$ by
\begin{equation}
\hat{\omega}^{ab} = \omega^{ab} - \eta^{ab} \phi\,.
\end{equation}
Note that $\hat{\omega}^{ab}$ is no more antisymmetric, and hence
does not take values in the Lorentz algebra so$(2,1)$ $\cong$ sl$(2,\bR)$,
but in gl$(2,\bR)$. Therefore, this connection is not metric, but it is
torsionless due to eq.~(\ref{e}),
\begin{displaymath}
d e^a + {\hat\omega^a}_{\;\;b} \wedge e^b = 0\,.
\end{displaymath}
This can be solved to give
\begin{eqnarray}
{\hat\omega}^{ab}_{\mu} &=& \frac 12 e^{a\nu}(\partial_{\mu} {e^b}_{\nu} - \partial_{\nu} {e^b}_{\mu})
                            + \frac 12 e^{a\nu} e^{b\lambda}(\partial_{\lambda} {e^c}_{\nu} -
                            \partial_{\nu} {e^c}_{\lambda}) e_{c\mu}\,, \nonumber \\
                        & & -\frac 12 e^{b\nu}(\partial_{\mu} {e^a}_{\nu} - \partial_{\nu} {e^a}_{\mu})
                            + e^{a\nu} \phi_{\nu}\,{e^b}_{\mu} - e^{b\nu} \phi_{\nu}\,{e^a}_{\mu}
                            - \eta^{ab} \phi_{\mu}\,.
\end{eqnarray}
As before, we require $\nabla_{\mu} {e^a}_{\nu} = 0$, which yields
\begin{eqnarray}
{\Gamma^{\mu}}_{\nu\rho} &=& {e_a}^{\mu} {\hat\omega}^a_{\nu b}\,{e^b}_{\rho} + {e_a}^{\mu}\partial_{\nu}
                             {e^a}_{\rho} \nonumber \\
                         &=& \lcc\Gamma{}^{\mu}{}_{\nu\rho} + \phi^{\mu} g_{\nu\rho} - \phi_{\rho}\,
                             {\delta^{\mu}}_{\nu} - \phi_{\nu}\,{\delta^{\mu}}_{\rho}\,,
                             \label{conncoeff}
\end{eqnarray}
where $\lcc\Gamma{}^{\mu}{}_{\nu\rho}$ denotes the Christoffel connection.
Using (\ref{conncoeff}), one obtains for the nonmetricity
\begin{equation}
\nabla_{\mu} g_{\nu\rho} = 2 g_{\nu\rho}\,\phi_{\mu}\,. \label{Weylconn}
\end{equation}
This is precisely the definition of a Weyl connection. Mathematically, a Weyl structure
on a manifold $\cal M$ is defined by a pair $W = (g, \phi)$, where $g$ and $\phi$ are
a Riemannian metric and a one-form on $\cal M$, respectively. There exists then one
and only one torsion-free connection $\nabla$, called the Weyl connection, such that
(\ref{Weylconn}) holds. Note that eq.~(\ref{Weylconn}) expresses the compatibility of
$\nabla$ with the conformal class of $g$. It is invariant under Weyl transformations
\begin{equation}
g_{\mu\nu} \to e^{2\chi} g_{\mu\nu}\,, \qquad \phi_{\mu} \to \phi_{\mu} + \partial_{\mu} \chi\,,
\end{equation}
where $\chi \in C^{\infty}({\cal M})$. Historically, the connection satisfying
(\ref{Weylconn}) was introduced by Weyl in 1919 in an attempt to unify general relativity
with electromagnetism \cite{Weyl:1919fi}.

We still have to interpret the equations (\ref{omega}), (\ref{lambda}) and (\ref{phi}).
(\ref{phi}) implies that the antisymmetric part of $\lambda_{\mu\nu}$ represents essentially
the field strength of $\phi$,
\begin{equation}
\lambda_{[\mu\nu]} = -\frac 12 (\partial_{\mu} \phi_{\nu} - \partial_{\nu} \phi_{\mu})
                     \equiv -\frac 12 f_{\mu\nu}\,.
\end{equation}
Let us denote the curvature two-form of the (metric, but not torsionless) connection
$\omega$ by $\cal R$, i.~e.~, ${\cal R}^{ab} = d\omega^{ab} + {\omega^a}_c \wedge \omega^{cb}$.
Note that the curvature of $\hat\omega$ splits as
\begin{displaymath}
R^{ab} = d{\hat\omega}^{ab} + {\hat\omega}^a_{\;\;c} \wedge {\hat\omega}^{cb}
       = {\cal R}^{ab} - \eta^{ab} f\,.
\end{displaymath}
Eq.~(\ref{omega}) is then equivalent to
\begin{equation}
{\cal R}_{\mu\nu\rho\sigma} = g_{\mu\rho} \lambda_{\nu\sigma} + g_{\nu\sigma} \lambda_{\mu\rho} -
                              g_{\mu\sigma} \lambda_{\nu\rho} - g_{\nu\rho} \lambda_{\mu\sigma}\,,
                              \label{omega2}
\end{equation}
which yields
\begin{displaymath}
\lambda_{\mu\nu} = {\cal R}_{\mu\nu} - \frac 14 {\cal R}^{\rho}_{\;\;\rho}\,g_{\mu\nu}\,,
\end{displaymath}
with ${\cal R}_{\mu\nu}$ denoting the Ricci tensor. Thus, $\lambda_{\mu\nu}$ represents
again the Schouten tensor, but the one constructed from the connection $\omega$, which
differs from the full Schouten tensor associated to $\hat\omega$ by a piece proportional
to the field strength $f_{\mu\nu}$. Eq.~(\ref{omega2}) expresses just the fact that
in three dimensions the curvature is determined by the Schouten tensor alone. Note that
this is still true in presence of torsion, cf.~appendix \ref{decompcurv}. Notice also
that ${\cal R}_{\mu\nu\rho\sigma}$ is antisymmetric in its first two indices by virtue
of the metricity of $\omega$, but that
${\cal R}_{\mu\nu\rho\sigma} \neq {\cal R}_{\rho\sigma\mu\nu}$, because $\omega$ has
nonvanishing torsion. Therefore the Ricci tensor ${\cal R}_{\mu\nu}$ is in general
not symmetric.

Finally, we come to eq.~(\ref{lambda}). Using $\nabla_{\mu} e_{a\nu} = 2\phi_{\mu}e_{a\nu}$,
one gets
\begin{equation}
\nabla_{\mu} \lambda_{\nu\rho} - \nabla_{\rho} \lambda_{\nu\mu} = 0\,. \label{lambda1}
\end{equation}
As $\lambda_{\mu\nu}$ does not represent the full Schouten tensor constructed from the
connection $\hat\omega$, this equation can not be interpreted as the vanishing of
the Cotton two-form associated to the Weyl connection. If one so wishes, one can
express $\lambda_{\mu\nu}$ in terms of the full Schouten tensor and the field strength $f$,
and use the Bianchi identity for $f$ to rewrite (\ref{lambda1}) as
\begin{equation}
C_{\mu\rho\nu} = -\nabla_{\nu} f_{\mu\rho}\,, \label{C=nablaf}
\end{equation}
where $C_{\mu\rho\nu}$ denotes the Cotton two-form constructed from the Weyl connection
$\hat\omega$. For zero ``electromagnetic'' field $\phi$, (\ref{lambda1}) means that the
spacetime is conformally flat. Equation (\ref{C=nablaf}) resembles the relation
\begin{equation}
C_{\mu\rho\nu} = -\frac 14 g_{\nu\rho} \nabla^{\lambda}f_{\lambda\mu}
                 +\frac 14 g_{\nu\mu} \nabla^{\lambda}f_{\lambda\rho}
                 -\frac 32 \nabla_{\nu} f_{\mu\rho}\,, \label{EWconstraint}
\end{equation}
that holds for Einstein-Weyl spaces \cite{Pedersen:1993}, i.~e.~, manifolds with
a Weyl connection for which the symmetric part of the Ricci tensor is proportional
to the metric,
\begin{equation}
R_{(\mu\nu)} = \frac R3 g_{\mu\nu}\,.
\end{equation}
The meaning of (\ref{C=nablaf}) can be clarified using the
expression \cite{Pedersen:1993}
\begin{equation}
R_{\mu\nu} = {\tilde R}_{\mu\nu} - \nabla_{\mu} \phi_{\nu} + 2\nabla_{\nu} \phi_{\mu}
             - g_{\mu\nu} \phi_{\lambda}\phi^{\lambda} + \phi_{\mu}\phi_{\nu}
             + g_{\mu\nu} \nabla_{\lambda} \phi^{\lambda} \label{RRtilde}
\end{equation}
for the Ricci tensor $R_{\mu\nu}$ of the Weyl connection in terms of the Ricci
tensor ${\tilde R}_{\mu\nu}$ of the Levi-Civita connection (not to be confused
with ${\cal R}_{\mu\nu}$), the one-form $\phi$ and its derivatives. Plugging
(\ref{RRtilde}) into (\ref{C=nablaf}) yields
\begin{equation}
{\tilde\nabla}_{\mu} {\tilde L}_{\nu\rho} - {\tilde\nabla}_{\rho}
{\tilde L}_{\nu\mu} = 0\,, \label{nablaLtilde}
\end{equation}
where $\tilde\nabla$ denotes the Levi-Civita connection and
${\tilde L}_{\mu\nu} = {\tilde R}_{\mu\nu} - \frac{\tilde R}4 g_{\mu\nu}$.
Eq.~(\ref{nablaLtilde}) means that, in terms of Riemannian data, spacetime is
conformally flat, so that we have a Weyl structure defined on a conformally
flat manifold.

The above results can also be understood from the point of
view of gauge transformations: It is clear that, at least
for an invertible triad, our model must be
gauge-equivalent to conformal gravity, i.~e.~, to the theory with $\phi_{\mu} = 0$.
This means that there must be gauge transformations that take any solution of our
theory to a conformally flat metric. Under a general gauge transformation $g$
the connection changes according to
\begin{equation}
A' = g^{-1} A g + g^{-1} dg\,. \label{Aprimo}
\end{equation}
For a special conformal transformation, $g = \exp(-\sigma^a K_a)$, (\ref{Aprimo})
leads to
\begin{displaymath}
\phi'_{\mu} = \phi_{\mu} - {e^a}_{\mu} \sigma_a\,.
\end{displaymath}
This vanishes if we choose $\sigma_a = {e_a}^{\mu} \phi_{\mu}$, which is always
possible for an invertible triad.

We conclude this section by noting that the similarity of (\ref{C=nablaf})
with equation (\ref{EWconstraint}), valid for Einstein-Weyl spaces,
suggests that there might be a relationship between the Einstein-Weyl equations
and the Chern-Simons action for the conformal group. Note in this context that
it is not known if the Einstein-Weyl equations follow from an action principle,
but it was conjectured in \cite{Pedersen:1993} that if such an action exists,
it might be related to gravitational Chern-Simons forms. It would be interesting
to explore this direction further.

\section{Metric affine gravity from SL$(4,\bR)$ Chern-Simons theory}

In the preceeding section we saw how to get Weyl structures from Chern-Simons
theory. The Weyl connection is a particular case of a nonmetric connection,
where only the trace part is present. One might ask whether it is possible to
write a more general metric affine gravity model as a Chern-Simons theory. We will
shew in this section that this is indeed possible. As was explained in section
\ref{mag}, in metric affine gravity, the Poincar\'{e} group ISO$(2,1)$ is
replaced by the affine group A$(3,\bR)$ $\cong$ GL$(3,\bR)$ $\ltimes$ $\bR^3$.
In attempting to write a CS action for the affine group, one encounters
the problem that the Lie algebra a$(3,\bR)$ is neither reductive nor is it the double
extension of some reductive Lie algebra, and therefore it does not admit an
Ad-invariant, nondegenerate quadratic form\footnote{In fact it is straightforward to
show explicitely that any Ad-invariant quadratic form on a$(3,\bR)$ is necessarily
degenerate.}.
A way out of this is to embed the
affine group in some slightly larger group that is semisimple. The most obvious thing
one can do is to consider the group SL$(4,\bR)$, which contains A$(3,\bR)$.
Let us denote the generators of SL$(4,\bR)$ by $L_{AB}$, $A = 0,\ldots,3$,
satisfying $\eta^{AB}L_{AB} = 0$, with $(\eta^{AB}) = \mathrm{diag}(-1,1,1,1)$.
They obey the commutation relations
\begin{displaymath}
[L_{AB}, L_{CD}] = \eta_{AD} L_{CB} - \eta_{CB} L_{AD}\,.
\end{displaymath}
Now split the generators into $L_{ab}$, $L_{3a} \equiv P_a$ and $L_{a3} \equiv K_a$,
where $a = 0,1,2$. In this way one obtains
\begin{eqnarray}
[L_{ab}, L_{cd}] &=& \eta_{ad}\,L_{cb} - \eta_{cb}\,L_{ad}\,, \nonumber \\
{[}L_{ab}, P_c] &=& \eta_{ac} P_b\,, \qquad [L_{ab}, K_c] = -\eta_{bc} K_a\,, \nonumber \\
{[}K_a, P_b] &=& - L_{ab} - \eta^{cd} L_{cd}\,\eta_{ab}\,, \\
{[}P_a, P_b] &=& [K_a, K_b] = 0\,. \nonumber
\end{eqnarray}
We see that $L_{ab}$ and $P_c$ generate the subgroup A$(3,\bR)$. The chosen
decomposition corresponds to rewriting the algebra sl$(4,\bR)$ as the graded
algebra $a^{\star} = \bR^3 \oplus \mathrm{gl}(3,\bR) \oplus {\bR^3}^{\star}$.
Although this seemingly looks like a generalization of the conformal algebra,
with so$(2,1)$ replaced by gl$(3,\bR)$, one cannot identify the $K_a$ with
the generators of special conformal transformations. In fact, SL$(4,\bR)$
does not contain the conformal group SO$(3,2)$ as a subgroup\footnote{Cf.~footnote
11 of \cite{Hehl:1994ue}.}.

Since SL$(4,\bR)$ is simple, it possesses (up to normalization) a unique
gauge-invariant bilinearform, given by
\begin{eqnarray}
&\displaystyle\langle L_{ab}, L_{cd}\rangle = \eta_{ad}\eta_{bc}-\frac14\eta_{ab}\eta_{cd}\,,\qquad
&\langle P_{a}, K_{b}\rangle = \eta_{ab}\,, \nonumber \\
&{\vphantom{\displaystyle\frac14}}\langle L_{ab}, P_{c}\rangle = \langle L_{ab}, K_{c}\rangle = 0\,,\qquad
&\langle P_{a}, P_{b}\rangle = \langle K_{a}, K_{b}\rangle = 0\,.
\end{eqnarray}
Let us decompose the connection according to
\begin{equation}
A = \sigma^{ba} L_{ab} + e^a P_a + \lambda^a K_a\,, \label{decomp}
\end{equation}
where we wish to interpret $\sigma^{ab}$ as a gl$(3,\bR)$-valued connection
and $e^a$ as the dreibein. The physical significance of $\lambda^a$ will
become clear later. The generator of infinitesimal gauge transformations
is a Lie-algebra valued zero-form,
\begin{displaymath}
u = \tau^{ba} L_{ab} + \rho^a P_a + \varsigma^a K_a\,.
\end{displaymath}
The variation of the gauge field $A$ under a gauge
transformation generated by $u$ is
\begin{displaymath}
\delta A = -\dd u - [A, u]\,.
\end{displaymath}
This means that the component fields transform as
\begin{eqnarray}
\delta\sigma^{ba} &=& -d\tau^{ba}-(\sigma^{b}{}_{c}\tau^{ca}-\tau^{b}{}_{c}\sigma^{ca})
                      - (e^{b}\varsigma^{a}+e^{c}\varsigma_{c}\eta^{ba}) +
                      (\lambda^{a}\rho^{b}+\lambda^{c}\rho_{c}\eta^{ba})\,, \label{deltasigma} \\
\delta e^{a} &=& -d\rho^a + \tau^{a}{}_{b}e^{b}-\sigma^{a}{}_{b}\rho^{b}\,, \label{deltae} \\
\delta \lambda^{a} &=& -d\varsigma^a - \lambda^{b}\tau_{b}{}^{a}+\varsigma^{b}\sigma_{b}{}^{a}\,.
                       \label{deltalambda}
\end{eqnarray}
With (\ref{decomp}), the Chern-Simons action becomes
\begin{eqnarray}
I_{CS} &=& \int\left(\si^a{}_b\wedge d\si^b{}_a+\frac23\si^a{}_b\wedge\si^b{}_c\wedge\si^c{}_a
         -\frac14\si\wedge d\si \right. \nonumber \\
       & & \qquad \left. + e_a \wedge d\lambda^a + \lambda_a \wedge d e^a +
           2\lambda_a \wedge \sigma^{ab} \wedge e_b \right)\,,
           \label{CSSL4R}
\end{eqnarray}
with $\si\equiv\si^a{}_a$.
The equations of motion following from (\ref{CSSL4R}) read
\begin{eqnarray}
de^a + {\sigma^a}_b \wedge e^b &=& 0\,, \label{T=0} \\
d\lambda^a - {\sigma_b}^a \wedge \lambda^b &=& 0\,, \label{C=0} \\
d\sigma^{ab} + {\sigma^a}_c \wedge \sigma^{cb} &=& - e^a \wedge \lambda^b -
e^c \wedge \lambda_c\,\eta^{ab}\,. \label{curv}
\end{eqnarray}
Eq.~(\ref{T=0}) means that the torsion vanishes, $T^a = 0$. Defining
\begin{displaymath}
\omega^{ab} \equiv \sigma^{[ab]}\,, \qquad Q^{ab} \equiv 2\sigma^{(ab)}\,,
\end{displaymath}
one can use (\ref{T=0}) and (\ref{N}) to obtain
\begin{equation}
{\Gamma^{\rho}}_{\mu\nu} = {{\tilde \Gamma}^{\rho}}_{\mu\nu} + \frac 12({Q^{\rho}}_{\mu\nu}
+ {Q^{\rho}}_{\nu\mu} - Q_{\mu\;\;\nu}^{\;\;\rho})\,, \label{Gamma}
\end{equation}
where ${{\tilde \Gamma}^{\rho}}_{\mu\nu}$ denotes the Levi-Civita connection and
${Q^{\rho}}_{\mu\nu}$ is given by
\begin{displaymath}
{Q^{\rho}}_{\mu\nu} = {e_a}^{\rho} Q^a_{\mu\,c} {e^c}_{\nu}\,.
\end{displaymath}
One easily verifies that $Q_{\nu\rho\mu} = Q_{\mu\rho\nu}$, and thus
${\Gamma^{\rho}}_{\nu\mu} = {\Gamma^{\rho}}_{\mu\nu}$, which is of course a
consequence of vanishing torsion. Using (\ref{Gamma}), one obtains for the
covariant derivative of the metric,
\begin{displaymath}
\nabla_{\mu} g_{\nu\lambda} = -Q_{\lambda\mu\nu}\,,
\end{displaymath}
so that $Q_{\lambda\mu\nu}$ represents the nonmetricity. Note that we also have
$\nabla_{\mu} e_{a\nu} = -e_{b\nu} Q^b_{\mu\,a}$. Using this and the
definition $\lambda_{\mu\nu} = e_{a\mu}{\lambda^a}_{\nu}$, eq.~(\ref{C=0})
is seen to be equivalent to
\begin{equation}
\nabla_{\mu} \lambda_{\alpha\nu} - \nabla_{\nu} \lambda_{\alpha\mu} = 0\,,
\label{Cotton2form}
\end{equation}
whose deeper meaning will become clear below.

We finally come to eq.~(\ref{curv}). The curvature two-form is defined by
\begin{equation}
R^{ab} = d\sigma^{ab} + {\sigma^a}_c \wedge \sigma^{cb}\,.
\end{equation}
Notice that in metric affine gravity, both the antisymmetry in the first two
indices, and the block symmetry $R_{\alpha\beta\mu\nu} = R_{\mu\nu\alpha\beta}$ of the
curvature tensor are lost. This is the reason why one can define two different Ricci tensors $R_{\mu\nu}$ and $S_{\mu\nu}$
(cf.~appendix \ref{decompcurv}). (\ref{curv}) yields
\begin{equation}
R_{\alpha\beta\mu\nu} = - g_{\alpha\mu}\lambda_{\beta\nu} + g_{\alpha\nu}\lambda_{\beta\mu}
                        - g_{\alpha\beta}\lambda_{\mu\nu} + g_{\alpha\beta}\lambda_{\nu\mu}\,,
                        \label{Riemann}
\end{equation}
and thus
\begin{eqnarray}
R_{\mu\nu} &\equiv& {R^{\alpha}}_{\mu\alpha\nu} = \lambda_{\nu\mu} - 3\lambda_{\mu\nu}\,, \nonumber \\
S_{\mu\nu} &\equiv& R_{\mu\;\;\nu\alpha}^{\;\;\alpha} = -\lambda_{\nu\mu} + 2\lambda_{\mu\nu}
                                                      - g_{\mu\nu}{\lambda^{\rho}}_{\rho}\,, \label{RS}
\end{eqnarray}
from which we get
\begin{equation}
\lambda_{\mu\nu} = -(R_{\mu\nu} - \frac R4 g_{\mu\nu} + S_{\mu\nu} - \frac S4 g_{\mu\nu})\,.
\end{equation}
Therefore, $\lambda_{\mu\nu}$ represents the sum of the two Schouten tensors constructed from
$R_{\mu\nu}$ and $S_{\mu\nu}$. Eq.~(\ref{Cotton2form}) means then that the sum of the two
Cotton two-forms that one can construct, must vanish. (\ref{Cotton2form}) is thus a
direct generalization of the field equation of conformal gravity.
The antisymmetric part of (\ref{Riemann}) yields
\begin{equation}
R_{[\alpha\beta]\mu\nu} = \frac 12 (- g_{\alpha\mu}\lambda_{\beta\nu} + g_{\beta\mu}\lambda_{\alpha\nu}
                                    + g_{\alpha\nu}\lambda_{\beta\mu} - g_{\beta\nu}\lambda_{\alpha\mu})\,,
\end{equation}
which means that the antisymmetrized curvature is given in terms of the sum of the two
Schouten tensors alone. This is in fact nothing else than the irreducible decomposition of
$R_{[\alpha\beta]\mu\nu}$ under the Lorentz group (cf.~appendix \ref{decompcurv}), which
comes out here as a field equation.

We still have to interpret the symmetric part of (\ref{Riemann}),
\begin{eqnarray}
R_{(\alpha\beta)\mu\nu} &=& \frac 12 (\nabla_{\mu} Q_{\alpha\nu\beta} - \nabla_{\nu} Q_{\alpha\mu\beta}) \nonumber \\
                        &=& -\frac 12 (g_{\alpha\mu}\lambda_{\beta\nu} + g_{\beta\mu}\lambda_{\alpha\nu}
                                     - g_{\alpha\nu}\lambda_{\beta\mu} - g_{\beta\nu}\lambda_{\alpha\mu})
                            - g_{\alpha\beta}(\lambda_{\mu\nu} - \lambda_{\nu\mu})\,.
                            \label{decompsymm}
\end{eqnarray}
In arbitrary dimension, the irreducible decomposition of
$R_{(\alpha\beta)\mu\nu} \equiv Z_{\alpha\beta\mu\nu}$ under
the Lorentz group involves five pieces $^{(i)}Z_{\alpha\beta\mu\nu}$, $i = 1,\ldots, 5$, with
$^{(2)}Z$ vanishing identically in three dimensions \cite{Hehl:1994ue}. Comparing the remaining
four pieces given in appendix \ref{decompcurv} with the decomposition (\ref{decompsymm})
implied by the field equations, we get
\begin{eqnarray}
^{(1)}Z_{\alpha\beta\mu\nu} &=& 0\,, \nonumber \\
\Delta_{\mu\nu} &=& -\frac 53 \lambda_{[\mu\nu]}\,, \nonumber \\
\Xi_{\mu\nu} &=& \frac 12 g_{\mu\nu}{\lambda^{\rho}}_{\rho} - \frac 32 \lambda_{(\mu\nu)}\,, \nonumber \\
^{(4)}Z_{\alpha\beta\mu\nu} &=& -\frac 83 g_{\alpha\beta}\lambda_{[\mu\nu]}\,.
\end{eqnarray}
A priori, the symmetric part of the curvature has 18 independent components in three dimensions,
but the field equation (\ref{decompsymm}) tells us that seven of them must vanish
($^{(1)}Z = 0$), and that the remaining ones are determined completely by the antisymmetric
and the symmetric traceless part of the Schouten tensor $\lambda_{\mu\nu}$, that
determines also the antisymmetric part of the curvature. Eqns.~(\ref{Cotton2form})
and (\ref{decompsymm}) are the only remaining equations for the metric $g_{\mu\nu}$ and
the nonmetricity $Q_{\lambda\mu\nu}$.

\subsection{Simple solutions}

A simple solution of these equations can be obtained
by setting $Q_{\lambda\mu\nu} = 0$. Then the connection reduces to the Christoffel
connection. Furthermore, $R_{\mu\nu} = S_{\mu\nu}$, and $\lambda_{\mu\nu}$ becomes symmetric.
In this case, eq.~(\ref{decompsymm}) is satisfied iff
$\lambda_{\mu\nu} = \frac 13 {\lambda^{\rho}}_{\rho}\,g_{\mu\nu}$, which implies
\begin{equation}
R_{\mu\nu} = \frac R3 g_{\mu\nu}\,,
\end{equation}
i.~e.~, the manifold is Einstein. For vanishing nonmetricity, we recover
therefore general relativity. Note that the cosmological constant appears here
as an integration constant, and not as an input. Notice also that
eq.~(\ref{Cotton2form}) is then identically satisfied, since in three dimensions
every Einstein space is conformally flat, and thus the Cotton two-form vanishes.
The fact that the cosmological constant is no more an external input
can be seen also from a group-theoretic point of view. The assumption of vanishing nonmetricity
selects the Lorentz generators from the ${\rm gl}(3,\mathbb{R})$ generators, reducing therefore
${\rm gl}(3,\mathbb{R})$ to ${\rm so}(2,1)$. The equations of motion imply that $\lambda^{a}$
is proportional to the dreibein,
\eq
\lambda^{a}=\frac{\lambda}3 e^{a}\,,
\feq
with $\lambda \equiv {\lambda^{\rho}}_{\rho}$.
Introducing the Lorentz generators $J_{ab}=-2L_{[ab]}$, the connection becomes
\eq
A=\frac12\omega^{ab}J_{ab}+e^{a}\left(P_{a}+\frac{\lambda}3K_{a}\right)\,,
\feq
so that the new translation generators are given by
\eq
\Pi_a = P_a + \frac{\lambda}3K_a\,.
\feq
The $J_{ab}$ and $\Pi_a$ obey the algebra
\begin{eqnarray}
\left[J_{ab},J_{cd}\right] &=& \eta_{ad}J_{bc}+\eta_{bc}J_{ad}-\eta_{ac}J_{bd}-\eta_{bd}J_{ac}\,,
                               \nonumber \\
\left[J_{ab},\Pi_{c}\right] &=& \eta_{bc}\Pi_{a}-\eta_{ac}\Pi_{b}\,,\\
\left[\Pi_{a},\Pi_{b}\right] &=& \frac{\lambda}3J_{ab}\,. \nonumber
\end{eqnarray}
Depending on the sign of $\lambda$ this is the algebra ${\rm so}(3,1)$ ($\lambda<0$),
${\rm so}(2,2)$ ($\lambda>0$) or iso$(2,1)$ ($\lambda = 0$), generated by the isometries of de~Sitter,
anti-de~Sitter or Minkowski spacetimes respectively, and the cosmological constant is given by $\La=-\la/3$.
An interesting observation is that these solutions enjoy a duality symmetry exchanging $e^a$ and
$\lambda^a$ and relating large and small cosmological constants. To be more precise, if we
act with the discrete transformation $e^a\mapsto\la^a$, $\la^a\mapsto e^a$, we obtain a new Einstein space
solving the model with a cosmological constant $1/\La$.

As a slight generalization let us consider the case when the nonmetricity has only a trace
part, i.~e.
\begin{displaymath}
Q_{\lambda\mu\nu} = -2 g_{\lambda\nu}\phi_{\mu}\,.
\end{displaymath}
This leads to
\eq
\nabla_{\mu} g_{\nu\lambda} = 2 g_{\nu\lambda} \phi_{\mu}\,,
\feq
so that $\nabla$ is a Weyl connection. If we define $F = \dd\phi$, eq.~(\ref{decompsymm})
yields
\eq
g_{\alpha\beta} F_{\mu\nu} = \frac 12 (g_{\alpha\mu}\lambda_{\beta\nu} + g_{\beta\mu}\lambda_{\alpha\nu}
                             - g_{\alpha\nu}\lambda_{\beta\mu} - g_{\beta\nu}\lambda_{\alpha\mu})
                             + g_{\alpha\beta}(\lambda_{\mu\nu} - \lambda_{\nu\mu})\,. \label{gF}
\feq
Contracting with $g^{\beta\mu}$ and taking the symmetric part, one obtains
\eq
\lambda_{(\alpha\nu)} = \frac 13 \lambda g_{\alpha\nu}\,. \label{symmlambda}
\feq
Using this in (\ref{RS}) we get
\eq
R_{(\mu\nu)} = S_{(\mu\nu)} = -\frac 23 \lambda g_{\mu\nu}\,,
\feq
which means that we have an Einstein-Weyl structure (cf.~e.~g.~\cite{Dunajski:2000rf}).
The antisymmetric part yields
\eq
\lambda_{[\alpha\nu]} = \frac 27 F_{\alpha\nu}\,. \label{antisymmlambda}
\feq
Inserting (\ref{symmlambda}) and (\ref{antisymmlambda}) in (\ref{gF}) and contracting
with $g^{\alpha\beta}$ leads to $F = 0$, so that $\phi$ is pure gauge, $\phi = \dd\chi$
locally. This pure gauge nonmetricity can be eliminated by conformally rescaling
\begin{displaymath}
g_{\mu\nu} \to {\hat g}_{\mu\nu} = e^{-2\chi} g_{\mu\nu}\,.
\end{displaymath}
The new metric $\hat g$ satisfies then the same equations as in the case $Q = 0$,
i.~e.~, it is an Einstein metric. $g$ is thus conformally Einstein.

\subsection{Partial gauge fixing}

As was explained in section \ref{mag}, the symmetry group that is gauged in
metric affine gravity is the affine group A$(3,\bR)$. On the other hand, our model
has the larger symmetry group SL$(4,\bR)$. In order to interpret the Chern-Simons
theory considered above as a model of metric affine gravity, we have to gauge fix
the additional symmetries, in the same way in which Horne and Witten gauge fixed
the special conformal symmetries of the SO$(3,2)$ CS theory considered in the
previous section. We first show that one can use the additional symmetries
generated by the $K_a$ to set the trace part of the connection $\sigma^{ab}$ to
zero. (\ref{deltasigma}) yields for the variation of the trace part under a
gauge transformation
\begin{equation}
\delta(\sigma^{ba}\eta_{ba}) = -d(\tau^{ba}\eta_{ba}) - 4\varsigma^a e_a
                               + 4\lambda^a \rho_a\,. \label{deltatrace}
\end{equation}
This shows that for an invertible triad, the $\varsigma^a$ gauge invariance
is precisely sufficient to set $\sigma^{ba}\eta_{ba} = 0$. Furthermore,
eq.~(\ref{deltae}) says that the triad is completely unchanged by a
$\varsigma^a$ gauge transformation, so ${e^a}_{\mu}$ remains invertible in this
new gauge. The gauge transformations
that preserve this gauge are given by the A$(3,\bR)$ generators $\tau^{ab}$ and
$\rho^a$, but from (\ref{deltatrace}) we see that we must compensate with a
$\varsigma^a$ transformation that is determined entirely by the $\tau^{ab}$ and
the $\rho^a$ according to
\begin{equation}
\varsigma_c = -\frac 14 {e_c}^{\mu}\partial_{\mu}(\tau^{ab}\eta_{ab})
              + {e_c}^{\mu}{\lambda^a}_{\mu}\rho_a\,. \label{comptransf}
\end{equation}
Note that ${\sigma^a}_a = 0$ implies ${Q^{\nu}}_{\mu\nu} = 0$, and thus
by (\ref{decompsymm}) $\lambda_{[\mu\nu]} = 0$, so the tensor $\lambda$
is symmetric in this gauge.

Next we show that the symmetries of the gauge fixed model consist of diffeomorphisms
and local GL$(3,\bR)$ transformations, as it should be for metric affine gravity.
If we set $\rho^a = \varsigma^a = 0$ in (\ref{deltasigma}) - (\ref{deltalambda}),
the $\tau^{ab}$ give a local GL$(3,\bR)$ transformation. Local diffeomorphisms are
not apparent in the transformation laws. Under a diffeomorphism generated by
$-v^{\mu}$, the fields should transform as
\begin{eqnarray}
\tilde{\delta} {e^a}_{\mu} &=& - v^{\nu}(\partial_{\nu} {e^a}_{\mu} - \partial_{\mu}
                              {e^a}_{\nu}) - \partial_{\mu} (v^{\nu}{e^a}_{\nu})\,,
                              \nonumber \\
\tilde{\delta}\sigma^{ab}_{\mu} &=& - v^{\nu}(\partial_{\nu} \sigma^{ab}_{\mu} -
                                    \partial_{\mu} \sigma^{ab}_{\nu}) -
                                    \partial_{\mu} (v^{\nu}\sigma^{ab}_{\nu})\,, \\
\tilde{\delta} {\lambda^a}_{\mu} &=& - v^{\nu}(\partial_{\nu} {\lambda^a}_{\mu} -
                                     \partial_{\mu}{\lambda^a}_{\nu}) -
                                     \partial_{\mu} (v^{\nu}{\lambda^a}_{\nu})\,. \nonumber
\end{eqnarray}
This should be a gauge transformation in our theory. If we make a gauge transformation
with gauge parameters $\rho^a = v^{\nu} {e^a}_{\nu}$,
$\tau^{ab} = v^{\nu}\sigma^{ab}_{\nu}$, $\varsigma^a = e^{a\mu}{\lambda^b}_{\mu}\rho_b$
(as required by (\ref{comptransf})), this differs from the diffeomorphism by
\begin{eqnarray}
\tilde{\delta} {e^a}_{\mu} - \delta {e^a}_{\mu} &=& -v^{\nu}(\partial_{\nu}{e^a}_{\mu}
- \partial_{\mu}{e^a}_{\nu} + \sigma^a_{\nu\,c} {e^c}_{\mu} -
\sigma^a_{\mu\,c} {e^c}_{\nu})\,, \nonumber \\
\tilde{\delta}\sigma^{ab}_{\mu} - \delta\sigma^{ab}_{\mu} &=& -v^{\nu}\left(\partial_{\nu}
\sigma^{ab}_{\mu} - \partial_{\mu} \sigma^{ab}_{\nu} - e^{b\rho}{\lambda^d}_{\rho}
e_{d\nu}{e^a}_{\mu} - {\lambda^d}_{\mu} e_{d\nu}\eta^{ab}\right. \nonumber \\
& & \left.\qquad + {\lambda^b}_{\mu}{e^a}_{\nu} + \lambda_{c\mu}{e^c}_{\nu}\eta^{ab}
- \sigma^a_{\mu\,d} \sigma^{db}_{\nu} + \sigma^{db}_{\mu} \sigma^a_{\nu\,d}\right)\,, \\
\tilde{\delta} {\lambda^a}_{\mu} - \delta {\lambda^a}_{\mu} &=& -v^{\nu}(\partial_{\nu}
{\lambda^a}_{\mu} - \partial_{\mu}{\lambda^a}_{\nu} + \sigma^{ba}_{\mu} {e_b}^{\rho}
{\lambda^c}_{\rho} e_{c\nu} - \lambda_{d\mu}\sigma^{da}_{\nu})\,. \nonumber
\end{eqnarray}
These differences vanish when the equations of motion (\ref{T=0}), (\ref{C=0})
and (\ref{curv}) hold, and when $\lambda_{\mu\nu} = \lambda_{\nu\mu}$, which is
satisfied in the gauge ${\sigma^a}_a = 0$ that we use. Thus, diffeomorphisms are
gauge transformations on shell.

\section{Conclusions}

It is possible to geometrically extend general relativity in several ways, by allowing torsion or
nonmetricity in the theory. In this article, we focused on three spacetime dimensions and showed how
to write such generalized gravitational models as Chern-Simons theories. Starting from the usual
formulation of three-dimensional gravity and using a non-standard decomposition of the Chern-Simons
connection, we recovered the Mielke-Baekler model for arbitrary sign of the effective cosmological
constant by playing with the independent coupling constants admitted by the gauge group.
In such a way, we realized explicitly three-dimensional gravity with torsion as a Chern-Simons theory.
Then, we turned to torsionless but nonmetric gravitational models. The simplest example is obtained by
allowing only the trace part of the nonmetricity. This is Weyl's gravity, and we proved its equivalence
with the SO$(3,2)$ Chern-Simons theory describing conformal gravity. Finally, we obtained a gravitational
theory with more general nonmetricity by embedding the affine group A$(3,\mathbb{R})$ in the special linear
group SL$(4,\mathbb{R})$ and writing a Chern-Simons action for the latter. It would be interesting to see
whether it is possible to obtain a gravitational theory incorporating both nonmetricity and torsion from a
Chern-Simons theory.

These gravitational models in reduced dimensionality are interesting because their integrability allows
to investigate important theoretical questions linked to the gravitational force. For instance,
we already mentioned the asymptotic dynamics of the MB model in the $\La_{\mathrm{eff}}<0$ case, which deserves
further analysis to identify the corresponding dual field theory. This, in turn, would give the opportunity
to understand the statistical mechanics of the Riemann-Cartan black hole.

Another important issue in theories where torsion and/or nonmetricity are present, is the coupling with
external matter. This is particularly problematic with generic nonmetricity, since the concept of light-cone,
and hence of causality, ceases to be invariant. However, in the Chern-Simons models of gravity under
consideration, one can easily write down an invariant action for a particle propagating on the backgrounds
they generate as a Wess-Zumino functional \cite{Skagerstam:1989ti},
\begin{equation}
S_{\mathrm p}=\int\!\!\dd\tau\left\langle K,g^{-1}\partial_\tau g\right\rangle\,,
\end{equation}
where $K$ is a constant element of the algebra, encoding the geometric properties of the particle (mass,
spin, etc.~) and $g(\tau)$ is an orbit of the gauge group of the gravitational theory under consideration. This
formulation has the advantage to provide straightforwardly a symplectic form for the Hamiltonian description
of the theory, which can be used to quantize the particle in a coordinate independent way. The analysis of
such systems would allow to define new intrinsic properties of the particles, analogous to mass and spin,
but corresponding to the additional generators of the gauge group. Hopefully, this could provide
some insight into the interpretation of metric affine theories, even in higher dimensional cases.

\acknowledgments

This work was partially supported by INFN, MURST and
by the European Commission program
MRTN-CT-2004-005104. We are grateful to F.~W.~Hehl for
clarifying correspondence and to G.~Ortenzi for discussions.
\normalsize

\appendix

\section{Decomposition of curvature in metric affine gravity}
\label{decompcurv}

In this section we briefly summarize the irreducible decomposition of
the curvature under the Lorentz group, given in \cite{Hehl:1994ue}.
Thereby, we specialize to three dimensions.

Let us first consider the antisymmetric part of the curvature. One easily
shows that
\begin{equation}
{\epsilon_{\gamma}}^{\alpha\beta} R_{\alpha\beta\mu\nu} {\epsilon^{\mu\nu}}_{\rho}
= 2 R_{\rho\gamma} - R\,g_{\rho\gamma} + 2 S_{\rho\gamma} - S\,g_{\rho\gamma}\,,
\label{epsReps}
\end{equation}
where
\begin{equation}
R_{\mu\nu} = {R^{\alpha}}_{\mu\alpha\nu}\,, \qquad
S_{\mu\nu} = R_{\mu\;\;\nu\alpha}^{\;\;\alpha}\,, \qquad R = {R^{\mu}}_{\mu}\,,
\qquad S = {S^{\mu}}_{\mu} = R\,.
\end{equation}
Note that in metric affine gravity, the Riemann tensor is no more symmetric in
the first two indices, so that one can define two different Ricci tensors
$R_{\mu\nu}$ and $S_{\mu\nu}$, that in general are not symmetric.
(For vanishing nonmetricity, but nonzero torsion, one has $R_{(\alpha\beta)\mu\nu} = 0$,
so that the two Ricci tensors coincide. However, since
$R_{\alpha\beta\mu\nu} \neq R_{\mu\nu\alpha\beta}$, the Ricci tensor is not
symmetric). Contracting (\ref{epsReps}) with
${\epsilon_{\lambda\sigma}}^{\gamma}{\epsilon^{\rho}}_{\tau\eta}$ yields
\begin{equation}
R_{[\sigma\lambda]\eta\tau} = \frac 12(g_{\sigma\eta} L_{\lambda\tau} +
g_{\lambda\tau} L_{\sigma\eta} - g_{\sigma\tau} L_{\lambda\eta} - g_{\lambda\eta}
L_{\sigma\tau})\,,
\label{decompanti}
\end{equation}
where $L_{\mu\nu}$ denotes the sum of the two Schouten tensors built from
$R_{\mu\nu}$ and $S_{\mu\nu}$,
\begin{equation}
L_{\mu\nu} = R_{\mu\nu} - \frac R4 g_{\mu\nu} + S_{\mu\nu} - \frac S4 g_{\mu\nu}\,.
\end{equation}
One can of course further decompose (\ref{decompanti}) into three pieces
corresponding to the antisymmetric, symmetric trace-free, and trace part of
the sum of the two Ricci tensors \cite{Hehl:1994ue}.

In order to decompose the symmetric part
$R_{(\alpha\beta)\mu\nu} \equiv Z_{\alpha\beta\mu\nu}$ of the curvature, one first
splits $Z$ into a traceless and a trace part,
\begin{displaymath}
Z_{\alpha\beta\mu\nu} = z_{\alpha\beta\mu\nu} + \frac 13 g_{\alpha\beta} {Z^{\gamma}}_{\gamma\mu\nu}\,.
\end{displaymath}
Then one gets \cite{Hehl:1994ue}
\begin{equation}
Z_{\alpha\beta\mu\nu} = {^{(1)}Z_{\alpha\beta\mu\nu}} + {^{(2)}Z_{\alpha\beta\mu\nu}} +
{^{(3)}Z_{\alpha\beta\mu\nu}} + {^{(4)}Z_{\alpha\beta\mu\nu}} + {^{(5)}Z_{\alpha\beta\mu\nu}}\,,
\label{decompZ}
\end{equation}
with $^{(2)}Z$ identically vanishing in three dimensions and
\begin{eqnarray}
^{(3)}Z_{\alpha\beta\mu\nu} &=& \frac 3{10}(g_{\alpha\mu}\,\Delta_{\beta\nu} -
g_{\alpha\nu}\,\Delta_{\beta\mu} + g_{\beta\mu}\,\Delta_{\alpha\nu} - g_{\beta\nu}\,
\Delta_{\alpha\mu}) - \frac 25 g_{\alpha\beta}\,\Delta_{\mu\nu}\,, \nonumber \\
^{(4)}Z_{\alpha\beta\mu\nu} &=& \frac 13 g_{\alpha\beta} {Z^{\gamma}}_{\gamma\mu\nu}\,, \nonumber \\
^{(5)}Z_{\alpha\beta\mu\nu} &=& \frac 13(g_{\alpha\mu}\,\Xi_{\beta\nu} -
g_{\alpha\nu}\,\Xi_{\beta\mu} + g_{\beta\mu}\,\Xi_{\alpha\nu} - g_{\beta\nu}\,\Xi_{\alpha\mu})\,,
\nonumber \\
^{(1)}Z_{\alpha\beta\mu\nu} &=& Z_{\alpha\beta\mu\nu} - {^{(3)}Z_{\alpha\beta\mu\nu}} -
{^{(4)}Z_{\alpha\beta\mu\nu}} - {^{(5)}Z_{\alpha\beta\mu\nu}}\,,
\end{eqnarray}
where
\begin{displaymath}
\Delta_{\mu\nu} = z_{\mu\;\;\alpha\nu}^{\;\;\alpha} - z_{\nu\;\;\alpha\mu}^{\;\;\alpha}\,,
\qquad
\Xi_{\mu\nu} = \frac 12 (z_{\mu\;\;\alpha\nu}^{\;\;\alpha} + z_{\nu\;\;\alpha\mu}^{\;\;\alpha})\,.
\end{displaymath}
In three dimensions, $Z_{\alpha\beta\mu\nu}$ has 18 independent components,
and (\ref{decompZ}) corresponds to the decomposition $18 = 7 + 0 + 3 + 3 + 5$.

\end{document}